\documentstyle[aps,epsfig]{revtex}
\tighten
\begin{document}
\draft
%__________________________________________________
%\twocolumn
%_______________________ Title, Authors ___________
%
\preprint{Preprint Numbers: \parbox[t]{45mm}{nucl-th/0112015}}
\title{Chirally symmetric quark description of low energy $\pi -\pi$ scattering}
\author{
Pedro Bicudo$^1$,
Stephen Cotanch$^2$,
Felipe Llanes-Estrada$^3$,
Pieter Maris$^2$,
Emilio Ribeiro$^1$, and
Adam Szczepaniak$^4$}
\address{
$^1$ Departamento de F\'\i sica, Instituto Superior T\'ecnico,
Av. Rovisco Pais, 1049-001 Lisboa, Portugal
\\
$^2$ Department of Physics, North Carolina State
University, Raleigh, NC 27695-8202
\\
$^3$  Departamento de
F\'{\i}sica Te\'orica I, Univ. Complutense de Madrid, 28040 Madrid,
Spain
\\
$^4$ Physics Department and Nuclear Theory Center
Indiana University, Bloomington, Indiana 47405-4202
}
\date{\today}
\maketitle
\begin{abstract}
Weinberg's theorem for $\pi -\pi$ scattering, including the Adler zero at threshold in 
the chiral limit, is analytically
proved for microscopic quark models that preserve chiral symmetry.
Implementing Ward--Takahashi  identities, the
isospin $0$ and $2$ scattering lengths are derived in exact agreement with Weinberg's
low energy results. Our proof applies to alternative quark formulations including
the Hamiltonian and Euclidean space Dyson--Schwinger approaches.   
Finally, the threshold $\pi -\pi$ scattering amplitudes are calculated using the  Dyson--Schwinger
equations in the rainbow-ladder truncation, confirming the formal derivation.
\end{abstract}
%%%%%%%%%%%%%%%%%%%%%%%%%%%%%%%%%%%%%%%%%%%%%%%%%%%%%%%%%%%%%%%%%%%%%%%%%%%%%
%
% 11.10.St      Bound and unstable states, BSEs
% 11.30.Rd      Chiral symmetries
% 12.38.Lg      Other non.pert. calculations
% 12.39.Pn      Potential models
% 12.40.Yx      Hadron mass models and calculations
% 13.20.-v      Leptonic and semileptonic decays of mesons
% 13.40.Gp      Electromagnetic form factors
% 13.75.Lb      Meson-meson interactions
% 14.40.-n      Mesons
% 14.40.Aq      pi K and eta mesons
% 14.40.Cs      Other mesons, S=C=0, mass< 2.5 GeV
% 24.85.+p      Quarks, gluons, and QCD in nuclei and nuclear processes
%
%%%%%%%%%%%%%%%%%%%%%%%%%%%%%%%%%%%%%%%%%%%%%%%%%%%%%%%%%%%%%%%%%%%%%%%%%%%%%
\section{Introduction}

In the zero quark mass limit the strong interaction is chirally
symmetric, exhibiting invariance under independent rotations of left
and right handed flavors.  Nonvanishing, albeit small on the hadronic
scale, $u$ and $d$ quark masses explicitly break this symmetry.  More
importantly, however, chiral symmetry is also broken spontaneously by
the vacuum and the corresponding Goldstone boson is the pion.  Thus,
it is generally accepted that chiral symmetry underlines the dynamics
of soft pions.  In particular, the low energy $\pi -\pi$ scattering
lengths can be calculated using current algebra and PCAC.  In the
seminal paper, based on these constraints, Weinberg~\cite{Weinberg}
deduced the scattering lengths for total isospin 0 and 2 to be
$a_0=+7L/4$ and $a_2=-L/2$, respectively, where $L=m^2_\pi /(8\pi
f^2_\pi)$ is given in terms of the pion mass, $m_\pi$, and decay constant,
$f_\pi$.
Further, chiral symmetry can also be used to constrain low energy,
effective theories describing pion-hadron interactions.  For example
in the linear $\sigma$ model, chiral symmetry enforces delicate
cancellations between direct $\pi -\pi$ interactions and
$\sigma$-exchange contributions to the scattering amplitude.

Concurrently, the advent of QCD has spawned the development
of more fundamental, microscopic formulations with quark degrees of
freedom and low energy hadronic phenomena has been successfully
described by various constituent quark models.  However, because
constituent quark models generally do not respect chiral symmetry, the
Goldstone nature of the pion is lost and there is no fundamental
difference between pion and, for example, the $\rho$ meson.
Consequently, such models should not be expected to properly describe
low energy pion scattering.  

It is, however, possible to construct quark models implementing chiral
symmetry, such as the rainbow-ladder truncation of the set of
Dyson--Schwinger equations [DSE] and the instantaneous Hamiltonian
approach using the random phase approximation [RPA], which 
preserve the pion's Goldstone nature.  For such
models it is remarkable that both calculated
observables and  attending mathematical relationships 
governed by chiral symmetry are
largely model independent,
even though 
gauge dependent nonobservable constructs, such as quark and gluon propagators,
can be very model sensitive.
The pion mass is a quintessential example
since for massless quarks this mass is zero, regardless of the form of
the effective interaction {\it provided} that there is spontaneous dynamical
chiral symmetry breaking ($S\chi SB$).  Similarly,  
$\pi$-$\pi$ scattering near threshold is governed by this symmetry and any
model that preserves chirality should reproduce Weinberg's scattering lengths.

The purpose of this paper is to demonstrate that for a microscopic
quark formulation with an arbitrary, but chiral symmetry preserving,
quark-antiquark interaction Weinberg's results can indeed be
explicitly obtained, and that a correct description of low-energy
$\pi$-$\pi$ scattering emerges.  The crucial step is to realize that
there are important higher order contributions to $\pi$-$\pi$ scattering 
which, in the chiral limit, must exactly cancel
the impulse contribution to recover the Adler zero.  Similar effects
occur in the microscopic Nambu--Jona-Lasinio model with contact
interactions at the quark level~\cite{Veronique}.  This is closely
related to the above mentioned cancellation in the $\sigma$ model.
The same results can also be derived in models using bosonization
techniques for the chiral pion fields as demonstrated in
Ref.~\cite{cr}.

This paper is organized into five sections.  In section II we utilize the
Hamiltonian formulation to 
succintly  derive Weinberg's results 
for the special case of an infinite interaction. 
Sections  III and IV detail the more general proof but now using the rainbow-ladder
truncation of the DSE.  We also demonstrate the direct
contribution to $\pi -\pi$ scattering 
violates chiral symmetry (does not vanish in the chiral limit) using
the axial-vector Ward--Takahashi identity 
[AV-WTI]
and provide a
numerical solution of the DSE further documenting the agreement with
Weinberg's scattering lengths. Finally, 
results and conclusions are summarized in
Sec.~\ref{secCon}.
%\nopagebreak [1]
%\samepage
%\newpage

%%%%%%%%%%%%%%%%%%%%%%%%%%%%%%%%%%%%%%%%%%%%%%%%%%%%%%%%%%%%%%%%%%%%%%%%%%%%%%
\section{Hamiltonian description of $\pi -\pi $ scattering}

In this section we provide a synopsis of our key result within the
Hamiltonian framework.  In section II.A we first address the one pion system and 
$S\protect\chi SB$.  Then in the following subsection we 
derive Weinberg's $\pi -\pi$ result in the
infinite interaction limit.

\subsection{Single pion formulation and $S\protect\chi SB$}

The Hamiltonian formalism for pions has an established history. 
For an earlier reference consult Ref. \cite{BDRELL} along with Refs.
\cite{origemdophi,peprd1992,fsprl,fs} for more recent applications. In the Hilbert space of pion
Salpeter amplitudes  the Hamiltonian can be expressed as
\begin{equation}
H=\sigma _{3}\ \left[ 
\begin{array}{c}
\Phi ^{+} \\ 
\Phi ^{-}
\end{array}
\right] \ m_{\pi }\ [\Phi ^{+},\Phi ^{-}]\ \sigma _{3}+\sigma _{3}\ \left[ 
\begin{array}{c}
\Phi ^{-} \\ 
\Phi ^{+}
\end{array}
\right] \ m_{\pi }\ [\Phi ^{-},\Phi ^{+}]\ \sigma _{3}  \label{eq:hampi}
\end{equation}
with 
\begin{equation}
H\ \left[ 
\begin{array}{c}
\Phi ^{+} \\ 
\Phi ^{-}
\end{array}
\right] =m_{\pi }\ \sigma _{3}\ \left[ 
\begin{array}{c}
\Phi ^{+} \\ 
\Phi ^{-}
\end{array}
\right] ;\ H\ \left[ 
\begin{array}{c}
\Phi ^{-} \\ 
\Phi ^{+}
\end{array}
\right] =-m_{\pi }\ \sigma _{3}\ \left[ 
\begin{array}{c}
\Phi ^{-} \\ 
\Phi ^{+}
\end{array}
\right].  \label{eq:HamilEIG}
\end{equation}
The normalization is given by 
\begin{equation}
\int d^3 k\lbrack \Phi ^{+},\Phi ^{-}]\ \sigma _{3}\ \left[ 
\begin{array}{c}
\Phi ^{+} \\ 
\Phi ^{-}
\end{array}
\right] =1  \label{eq:PINOR}
\end{equation}
where $\sigma _{3}$ is the standard third component Pauli matrix  and $\Phi
^{+},\Phi ^{-}$ represents the pion positive, negative energy components, respectively. 
For $%
\Phi ^{+}\rightarrow \Phi ^{-}$,  $m_{\pi }\rightarrow -m_{\pi }$ 
while for $\Phi ^{+}=\Phi ^{-}$, $m_{\pi }=0$. 
%The latter is the Goldstone
%pion and  in this particular case the pion wavefunction does not have a well defined normalization.

%Equations (\ref{eq:hampi}, \ref{eq:HamilEIG} and \ref{eq:PINOR}) together
%with the definition for $\Phi ^{\pm }$,
Previous pion Hamiltonian studies \cite{origemdophi,peprd1992,fsprl,fs} have utilized 
the Bogolubov-Valatin [BV] (or Bardeen, Cooper, Schriffer
[BCS]) transformation approach to describe the ground state
vacuum and developed quasiparticle (rotated quark/anti-quark
creation and annilhilation) operators for describing the
pion in the random phase approximation RPA \cite{fsprl,fs}.  The RPA
Hamiltonian formalism rigorously preserves chiral symmetry
and is formally equivalent to the rainbow-ladder DSE, Bethe--Salpeter equation [BSE]
approach with an
instantaneous kernel provided the quark propagator is
consistent with the BCS vacuum.  As detailed elsewhere \cite{Salpeters} the pion
momentum wavefunction can be expressed in terms of the BV rotation (or BCS
gap) angle $\phi(k)$    
\begin{equation}
\Phi ^{\pm }=\frac{\sin (\phi )}{a}\pm a\ \Delta ; \  \; \; a= \sqrt{\frac{2}{3%
}}\ f_{\pi }\ \sqrt{m_{\pi }} \ . \label{eq:PIDEFDELTA}
\end{equation}
For $m_{\pi }=0$, 
$\Delta =0$ and the rest frame pion Salpeter amplitudes reduce to $\Phi ^{\pm
}=\sin(\phi )/(0)$ which are degenerate and non-normalizable. 

The derivation of the above equations follows  
%Eqs.(\ref{eq:hampi}, \ref{eq:HamilEIG}, \ref{eq:PINOR}) follows 
straightforwardly  from an instantaneous reduction of
the pion Dyson--Schwinger, Bethe--Salpeter
equation  which is  depicted in  Fig. 1.
References \cite{Salpeters,ladders,NRDPH} contain a more complete derivation along with
applications.
In this formulation the mechanism of spontaneous chiral symmetry breaking is simple
to  understand: it corresponds to a self-consistent choice of the fermion Fock space appropriate 
to the quark kernel in use.  The Pauli principle still permits an infinite set 
of  Fock spaces which can be isomorphically  mapped to an infinite set 
of functions with coordinates $\phi (k) $. 
Any such Fock space can be obtained from the trivial one by a 
BV transformation with a $\phi (k)$ that is determined by solving the mass gap
equation.  This in turn specifies the  physical Fock space. For further
information, including the origin of $\phi$, consult Refs.
\cite{origemdophi,peprd1992,fsprl,fs}.

%%%%%%%%%%%%%%%%%%%%%%%%%%%%%%%%%%%%%%%%%%%%%%%%%%%%%%%%%%% 
\begin{figure}[htb]
\label{BSpi}
\epsfxsize=12cm
\begin{center}
\epsfbox{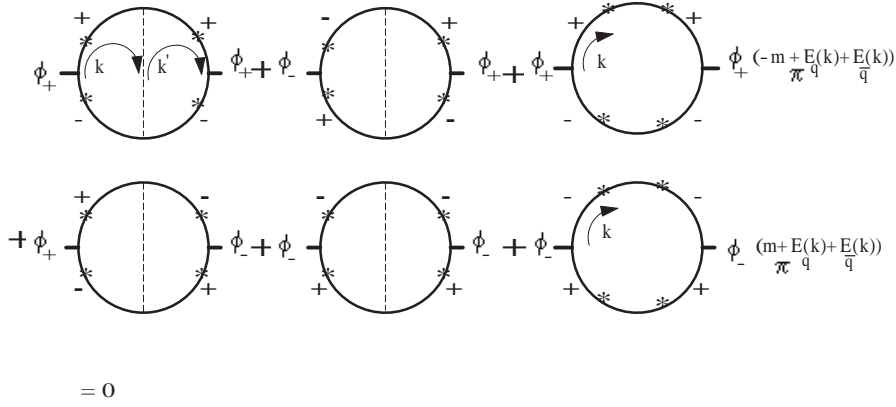}
%the above line generates a postscript error on my mac latex
\end{center}
\vspace{10mm}
\caption{ Pion Salpeter equation. 
In terms of the Dirac matrices $\beta$ and $\vec{ \alpha}$, the projection operators for the quark
propagator, with momentum $\vec {k}$, are 
$\Lambda^{\pm}=(1\pm\sin (\phi )\beta\pm\cos (\phi )
{\bf\alpha }\cdot \widehat{ {\bf k}})/2$, and denoted in the figure by $\{+,-\}$. 
Note that $\Phi^{\pm}$  is consistent
with the normalization  condition, Eq. (\ref{eq:PINOR}), and should 
contain the cluster propagators obtained after integrating the quark propagator energy, $E_q$. This 
is the reason the propagator cuts are displayed in the figure. Two such cluster propagators 
are needed for the two
$\Phi $' s but only one is generated per  integration loop.  This necessitates multiplying 
and dividing
the diagrams by the missing cluster propagator leading to the factors 
$\pm m_{\pi} +E_{q}+E_{\bar q}$ appearing in the  diagram.}
\end{figure}
%%%%%%%%%%%%%%%%%%%%%%%%%%%%%%%%%%%%%%%%%%%%%%%%%%%%%%%%%%%%

\subsection{$\pi -\pi $ scattering}

For $\ \pi -\pi $
scattering we can repeat the steps represented in Fig. 1.
This is diagrammatically summarized in Fig. 2.  
%Using Eqs. (\ref{eq:hampi}), (\ref{eq:HamilEIG}) and (\ref{eq:PINOR}) and
%the expression for $\Phi ^{\pm }$ we can immediately derive Weinberg's result
%for the limiting case of an infinite interaction.  
Here 
$\{ 1,2\}$ represent the  
incoming pions while $\{ 3,4\}$ denote the outgoing particles. 
Both the initial and final configurations now have two energy-spin Salpeter
amplitudes. In total  there are 48 diagrams plus kinetic energy insertions. The $\pm$
superscripts for pions 1...4 represent the different energy-spin amplitudes. 
Notice that these 
diagrams closely resemble those of Fig. 1 so that we could anticipate the 
model independence of $\pi -\pi $ scattering lengths since the map 
$\{\pi\times\pi\rightarrow \pi\}$ respects the structure of the pion 
Salpeter equation \cite{pipi}. 
%For a derivation of $\pi -\pi $ 
%scattering lengths in the framework of a contact Nambu, Jona-Lasinio 
%interaction see Ref.\cite{Veronique} as well as Ref.\cite{cr} which uses a
%global color-symmetry model.

\begin{figure}[htb]
\label{pipiscat}
\epsfxsize=12cm
\begin{center}
\epsfbox{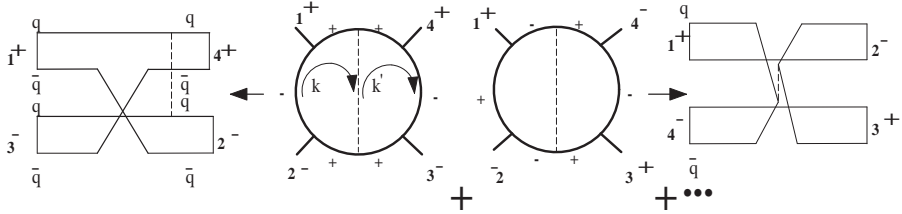}
\end{center}
\vspace{10mm}
\caption{Representative $ \pi\ -\pi$ scattering diagrams.}
\end{figure}

Figure 2 clearly indicates that both quark exchange {\em and} quark 
annihilation amplitudes are 
necessary. Of course this is a consequence of the quark Dirac nature, however,  
it is important to note that with just 
quark exchange only repulsion is obtained. This is 
precisely the case of exotic scattering. It is chiral symmetry which 
governs the correct combination of exchange and annihilation diagrams to 
yield a Goldstone pion and by doing so, to produce the Weinberg 
scattering lengths.

We now evaluate the isospin $ I = 0$ and $2$ $\pi -\pi $  $T$ matrix, $T_I$, or
scattering amplitude $A_I = i 4 \pi m_\pi T_I$, at zero pion
momentum.  Either from  comparing Figs. 1 and 2 or from Eqs. (\ref{eq:hampi}), 
(\ref{eq:HamilEIG}) and (\ref{eq:PINOR}) it is clear that   
$T_I$  will be proportional to $m_\pi $ because H is proportional to $m_\pi $.  
Further, $T_I$ is also inversely proportional to $f_\pi $ since there are {\em just} two extra
pion amplitudes when going from the Salpeter equation for one single
pion to
$\pi -\pi $ scattering. After  including all potential energy diagrams plus 
kinetic insertions we have,
\begin{equation}
\label{TMATRIX}
T_I=b_{I} \left[ {\Phi^{-}}^2,{\Phi^{+}}^2\right] H
\left[ 
 \begin{array}{c}
{\Phi^{+} }^2 \\ 
{\Phi^{-} }^2 \end{array}
 \right] 
+c_{I}\left[ \Phi^{+}\Phi^{-},\Phi^{+}\Phi^{-}\right] H
\left[ 
 \begin{array}{c}
\Phi^{+}\Phi^{-}\\ 
\Phi^{-}\Phi^{-}\end{array}
 \right]+
d_{I} \left[ {\Phi^{+}}^2,{\Phi^{-}}^2\right] H
\left[ 
 \begin{array}{c}
{\Phi^{+} }^2 \\ 
{\Phi^{-} }^2 \end{array}
 \right] \ .
\end{equation}
In the above expression integration is assumed. The Hamiltonian is given by
Eq. (\ref{eq:HamilEIG}) and the constants,  entailing color, spin and isospin traces, are 
\begin{equation}
b_{[I=0,I=2]}=\left[ 3/4,0\right];\;
c_{[0,2]}=\left[ 3/4,0\right];\;
d_{[0,2]}=\left[-1/2,1\right] \ .
\end{equation}

Equation (\ref{TMATRIX}) is especially simple to evaluate for an infinite interaction 
since  $\sin \phi (k)\rightarrow 1$. This limit always exists because 
for a given class of the quark kernels (for instance linear confinement with potential
$\sigma r$) the mass gap
equation for 
$\phi (k)$ can always be re-scaled as an dimensionless equation for $\phi (k/\sigma)$ for an
arbitrary  kernel strength $\sigma$. In the extremely strong limit, which also corresponds to
the pion point limit, 
$\sigma
\rightarrow
\infty$, and $\phi(k/\sigma) \rightarrow \phi(0) = \pi/2$  which is  {\em
sufficient} to obtain Weinberg's scattering lengths. To appreciate this use Eq.(\ref{eq:PIDEFDELTA})
repeatively with
$\sin(\phi) = 1$ to obtain,
\begin{equation}
\label{resint}
\int \left[  
{\Phi}^{+},{\Phi}^{-} 
\right] 
\;\sigma_{3}\;
\left[ 
 \begin{array}{c}
\Phi^{+} \\ 
\Phi^{-} \end{array}
 \right] =\int 4 \sin (\phi )\Delta=1;\;
\int \left[ 
\Phi^{-},\Phi^{+} 
\right]
\;\sigma_{3}\;
\left[ 
 \begin{array}{c}
{\Phi^{+} }^2 \\ 
{\Phi^{-} }^2 \end{array}
 \right]=\int 2 \sin^2 (\phi )\Delta=1/2 
\end{equation}
and so on...
It is then a text book calculation to extract, in lowest order in 
$m_\pi $, the $\pi -\pi $ scattering lengths from 
the $T$ matrix, Eq.(\ref{TMATRIX}), since in the Born approximation
$a_I= m_{\pi} T_I /(4 \pi )$ and we get 
the desired result $a_{[0,2]}=[7/4, -1/2] L$.
%%%%%%%%%%%%%%%%%%%%%%%%%%%%%%%%%%%%%%%%%%%%%%%%%%%%%%%%%%%%%%%%%%%%%%%%%%%%%

With this  preliminary treatment we now address  
the more general derivation for an arbitrary but finite quark kernel. First we
demonstrate that the impulse approximation is insufficient for $\pi -\pi$ scattering 
and violates chiral symmetry.

%%%%%%%%%%%%%%%%%%%%%%%%%%%%%%%%%%%%%%%%%%%%%%%%%%%%%%%%%%%%%%%%%%
\section{Impulse approximation for $\pi -\pi$ scattering}

\begin{figure}[h]
\center
\epsfig{figure=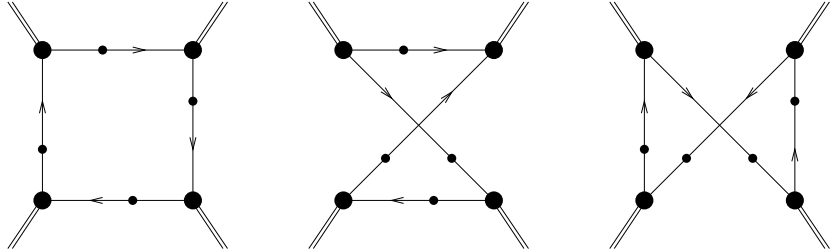,width=12cm}
\caption{Three different topologies for the direct contribution
to $\pi -\pi$ scattering.  Reversing the quark and anti-quark flow
(reversing the arrow)  generates three other diagrams for a total
of six.  The larger solid cirlces represent the pion-quark vertex while the smaller
solid circles denote the quark propagators are dressed.}
\label{figdirect}
\end{figure}

In this section we evaluate the leading (box) diagrams for $\pi -\pi$
scattering and demonstrate that they  fail
to provide a correct description since they violate chiral symmetry.  These diagrams, which we
call direct terms, correspond to the impulse approximation and are
illustrated in Fig.~\ref{figdirect}. 

We can evaluated the direct
terms model-independently using the AV-WTI which exactly relates the
axial-vector, $\Gamma_\mu^5$, and pseudoscalar, $\Gamma^5$, vertices
to the inverse of the dressed quark propagator
\begin{eqnarray}
  -i \left( P^\mu \Gamma_\mu^5(p',p;P)
        - 2 \, m_q(\mu) \,\Gamma^5(p',p;P) \right) & = &
	S^{-1}(p') \gamma_5 + \gamma_5 S^{-1}(p) \, .
\label{avwti}
\end{eqnarray}
Here $p$, $p'=p+P$ are the respective incoming, outgoing quark
momenta, and $P$ the momentum flowing into the vertex. The inverse of
the dressed quark propagator, $S^{-1}(p)$, can be expressed in terms
of scalar functions $A$ and $B$
\begin{eqnarray}
 i S^{-1}(p) & = & \, A(p^2) \;p\!\!\!/ - B(p^2) \, .
\end{eqnarray}
For the free propagator $A=1$ and $B=m_q(\mu)$, the current quark mass
which explicitly breaks chiral symmetry.  Since the interaction
contains a diverging, but renormalizable, short-range component, this mass requires
renormalization and depends on scale $\mu$.

Meson bound states can be described by a Bethe--Salpeter amplitude
[BSA] satisfying a homogeneous BSE
\begin{eqnarray}
  \Gamma^{ab}_H(p',p;P) & = & \int_k K^{ab;cd}(p',p;k,k') \;
                \left[S(k')\,\Gamma_H(k',k;P) \,S(k) \right]^{dc} \,,
\end{eqnarray}
with $\int_k$ representing $\int \frac{d^4k}{(2\pi)^4}$ and $K$ is the
quark-antiquark scattering kernel, properly regularized for divergent
integrals.  This equation only has solutions for discrete values of
$P^2$ corresponding to the bound state mass.  The lowest bound state in
the pseudoscalar channel is the pion, $P^2 = M^2_\pi$, which produces
poles in both $\Gamma_\mu^5$ and $\Gamma^5$.  For the axial-vector
vertex the residue of this pole is $f_\pi P_\mu$, while the residue of
pseudoscalar vertex is labeled $r_P(\mu)$. Both can be calculated from
the properly normalized pion BSA and the dressed quark propagators
\begin{eqnarray}
  P_\mu\, f_\pi  & = &  i \, Z_2 \int_k
	{\rm Tr}\left[ S(k') \; \Gamma_\pi(k',k;P) \; 
	S(k) \; \gamma_\mu \gamma_5 \right] \,,
\label{eqnfpi}
\\
  m_q(\mu) \; r_P(\mu) & = &  -Z_4
	\, m_q(\mu)\, \int_k {\rm Tr}\left[ S(k') \;   
	\Gamma_\pi(k',k;P) \; S(k) \; \gamma_5 \right] \,.
\label{eqnrps}
\end{eqnarray}
The constants $Z_2$ and $Z_4$ are the usual quark wave function and
mass renormalization terms, respectively.  They depend on both
renormalization and regulator scales but when combined with the
integrals and $m_q(\mu )$ incorporated in the above expressions, the
final results are scale independent~\cite{Maris:1998hd}.  This has
been checked explicitly for a model having an effective interaction
that reduces to the one-loop running coupling~\cite{Maris:1997tm}.
Finally, these residues are constrained to cancel through the AV-WTI
\begin{eqnarray}
  M^2_\pi \,f_\pi - 2 \; m_q(\mu) \; r_P(\mu) & = & 0 \, .
\label{avwtipoles}
\end{eqnarray}
Even though there is no pion pole in the inverse quark propagator, we
can still use the AV-WTI to express $\Gamma_\pi$ in terms of the
inverse quark propagator. In the combined chiral and $P_\mu\rightarrow
0$ limit this yields
\begin{eqnarray}
\Gamma_\pi(p,p;0) & = & \frac{B_0(p^2)}{f_\pi} \gamma_5
\label{avwti0chi}
\end{eqnarray}
where the subscript 0 indicates the chiral limit.

The evaluation of the direct terms
in Fig.~\ref{figdirect} produces
\begin{eqnarray}
 \int_k {\rm Tr}[\Gamma_\pi(k+P_1,k;P_1) \, S(k) \,
\Gamma_\pi(k,k-P_2;P_2) \, S(k-P_2)
\bar{\Gamma}_\pi(k-P_2,k-P_2-P_3;P_3) \,
\nonumber \\ {}\times
	S(k-P_2-P_3) \, \bar{\Gamma}_\pi(k-P_2-P_3,k+P_1;P_4)
        \, S(k+P_1) ] \;,
\end{eqnarray}
which at threshold and along with Eq. (\ref {avwti0chi}), reduces in the chiral limit to
\begin{eqnarray}
 \frac{1}{f_\pi^4} \int_k \frac{B_0^4(k)}{(k^2 A_0^2(k)
	+ B_0^2(k))^2} & \neq & 0  \;.
\end{eqnarray}
Note that this direct term result, which is exact since 
it follows from the AV-WTI, is not zero in the chiral limit.  From Weinberg's
theorem, however, the $\pi -\pi$ scattering amplitude at threshold
scales as $ M^2_\pi/{f^2_\pi}$ and therefore must vanish in the
chiral limit.  Clearly the direct term alone is insufficient to obtain
Weinberg's result and additional diagrams are necessary.

%%%%%%%%%%%%%%%%%%%%%%%%%%%%%%%%%%%%%%%%%%%%%%%%%%%%%%%%%%%%%%%%%%
\section{Dyson--Schwinger in the Rainbow-Ladder Approach}
\label{secDSE}

The problem raised in the previous section can be resolved by
utilizing a formalism which preserves chiral symmetry.  In this
section we consider one such approach, the Dyson--Schwinger method in
the rainbow ladder truncation (see Ref.\cite{robalk} for a review and applications).  The set of DSEs
form a hierarchy of coupled integral equations for the Green's function of the underlying
theory.  For example, the quark propagator $S(p)$ satisfies
\begin{eqnarray}
 iS^{-1}(p)
& = &  Z_2\;p\!\!\!/ - Z_4\,m_q(\mu)
        - i\int_k  g^2 \; \gamma_\mu \frac{\lambda^i}{2} \; S(k)
        \; \Gamma^i_\nu(k,p) \; D^{\mu\nu}(k-p) \,,
\label{quarkdse}
\end{eqnarray}
where $\Gamma^i_\nu(k,p)$ is the dressed quark-gluon vertex with gluon
color label $i = 1 \ldots 8$ and $D^{\mu\nu}(k-p)$ is the gluon
propagator.  The constants $Z_2$ and $Z_4$ follow from the
renormalization condition
\begin{eqnarray}
 i S^{-1}(p)\Big|_{p^2=\mu^2} & = &  p\!\!\!/ - m_q(\mu) 
\end{eqnarray}
at the renormalization scale $\mu$.  The equation for the axial-vector
vertex $\Gamma_\mu^5(p',p;P)$ and the pseudoscalar vertex
$\Gamma^5(p',p;P)$ is the generic inhomogeneous BSE
\begin{eqnarray}
  \Gamma^{ab}(p',p;P) & = &
\gamma_{\hbox{\scriptsize inhom}}
        + \int_k K^{ab;cd}(p',p;k,k') \;
\left[S(k')\,\Gamma(k',k;P)\,S(k)\right]^{dc} \,,
\end{eqnarray}
with the inhomogeneous terms $Z_2\,\gamma_\mu\,\gamma^5$ and
$Z_4\,\gamma^5$, respectively.  It is convenient to define an axial
vertex $\Gamma_A$
\begin{eqnarray}
  \Gamma_A(p',p;P) & = & -i \left( P^\mu \Gamma_\mu^5(p',p;P)       
	- 2 \, m_q(\mu) \,\Gamma^5(p',p;P) \right) \,,
\end{eqnarray}
which satisfies the BSE with inhomogeneous term
\begin{eqnarray}
  \gamma_A(P) & = & -i \left( Z_2 \, P\!\!\!\!/\,
	- 2 \, Z_4 \, m_q(\mu) \right) \gamma^5 \,.
\label{gammaAren}
\end{eqnarray}
For this vertex, the AV-WTI reduces to
\begin{eqnarray}
  \Gamma_A(p',p;P) & = & S^{-1}(p') \gamma_5 + \gamma_5 S^{-1}(p) \,.
\label{avwtiA}
\end{eqnarray}
From Eqs.~(\ref{eqnfpi})-(\ref{avwtipoles}) it follows that for any four-momentum 
$Q$ flowing into the
diagram we have
\begin{eqnarray}
  \int_k {\rm Tr}\left[ S(k+P) \; \Gamma_\pi(k+P,k;P)
\; S(k) \;
        \gamma_A(Q) \right] & = & (P \cdot Q + M^2_\pi)\, f_\pi
\,.
\label{pedro36}
\end{eqnarray}
Finally, expanding the vertex in powers of $P$, valid for low
momentum, small quark and pion masses, yields
\begin{eqnarray}
  \frac{i}{2\;f_\pi} \Gamma_A(p',p;P) & = &
	\Gamma_\pi(p',p;P) + {\cal O}(P) + {\cal O}(m_q(\mu))\, .
\label{expansion}
\end{eqnarray}
%%%%%%%%%%%%%%%%%%%%%%%%%%%%%%%%%%%%%%%%%%%%%%%%%%%%%%%%%%%%%%%%%%
\subsection{Rainbow-ladder truncation}

\begin{figure}[h]
\center
\epsfig{figure=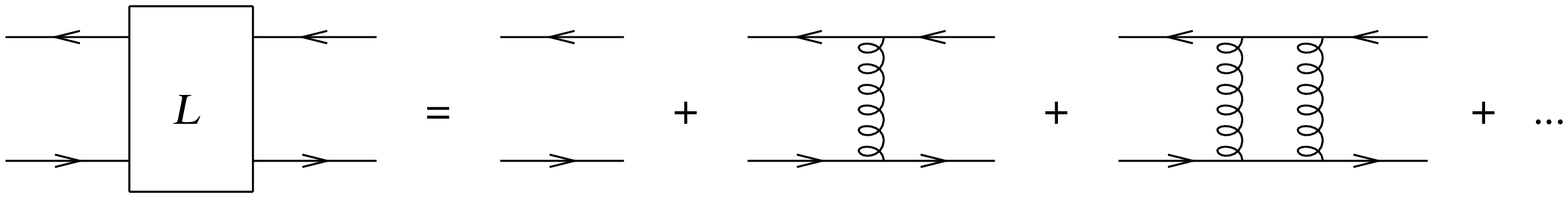,width=12cm}
\caption{
The unamputated quark-antiquark scattering amplitude $L$
in the ladder truncation.  The quark propagators are dressed, 
but for simplicity the solid circles indicating
the dressing are omitted in this and subsequent diagrams.}
\label{figladder}
\end{figure}

\begin{figure}[h]
\center
\epsfig{figure=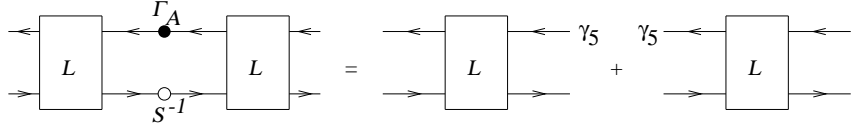,width=12cm}
\caption{
Reduction of connected ladder amplitudes.}
\label{figladderA}
\end{figure}

Provided that the regularization scheme is translationally invariant,
the rainbow truncation for the quark DSE
\begin{eqnarray}
  iS^{-1}(p) & = &  Z_2\;p\!\!\!/ - Z_4\,m_q(\mu)
        - i\int_k  \gamma_\mu \frac{\lambda^i}{2} \; S(k) \;
	\gamma_\nu \frac{\lambda^i}{2}\; g^2 D^{\mu\nu}(k-p) \,,
\label{quarkdseladder}
\end{eqnarray}
combined with the ladder truncation for the quark-antiquark scattering
kernel
\begin{eqnarray}
K^{ab;cd}(p',p;k,k') &=& \gamma_\mu^{ad}\frac{\lambda^i}{2} \;
	\gamma_\nu^{cb} \frac{\lambda^i}{2}\; g^2 D^{\mu\nu}(p-k) \,,
\label{kernel}
\end{eqnarray}
is consistent with the Ward identities\cite{Maris:1997tm,del}.  To discuss $\pi -\pi$
scattering, we introduce the unamputated quark-antiquark scattering
amplitude, $L$, in the ladder truncation which is pictorially
represented in Fig.~\ref{figladder}.  Here the solid horizontal lines
represent quark (antiquark) propagation and the coiled vertical lines
correspond to the quark-antiquark kernel, $K$, in the ladder
truncation, Eq. (\ref{kernel}).
This scattering amplitude satisfies the following DSE in the ladder truncation
\begin{eqnarray}
  L^{ab;cd}(p',p;k,k') & = &
        S^{ad}(p) \; S^{cb}(k) \; \delta^4(p-k)  \nonumber
 \\ & & {}
        -\int_q  S^{aa'}(p') \, \gamma_\mu^{a'd'} \,
        L^{d'c';cd}(q',q;k,k') \, \gamma_\nu^{c'b'} \, S^{b'b}(p)
        \, g^2 D^{\mu\nu}(p-q) \; .
\label{defL}
\end{eqnarray}
Using the AV-WTI, the ladder truncated amplitudes connected by the
vertex $\Gamma_A$, depicted in Fig.~\ref{figladderA}, can be reduced to
(see Ref.\cite{NRDPH} for a similar reduction of the vector vertex) 
\begin{eqnarray}
\lefteqn{ \int_q L^{ab;b'a'}(p'+Q,p;q,q'+Q) \;
                \Gamma_A^{a'd'}(q'+Q,q';Q) \;
                L^{d'c';cd}(q',q;k,k') \; (S^{-1})^{c'b'}(q) }
\nonumber \\
        & = & \gamma_5^{aa'} \; L^{a'b;cd}(p',p;k,k')  +
                L^{ab;cd'}(p'+Q,p;k,k') \; \gamma_5^{d'd} \, .
\label{emilio15}
\end{eqnarray}
Related, it can also be shown in the combined rainbow-ladder
truncation that
\begin{eqnarray}
\lefteqn{ \int_k L^{ab;cd}(p+P_1,p-P_2;k-P_2,k+P_1) \;
        \left[ \Gamma_A(k+P_1,k;P_1) \; S(k) \;
        \Gamma_A(k,k-P_2;P_2) \right]^{dc} }
\nonumber \\  & = &
  \left[\gamma_5\,S(p)\,\Gamma_A(p,p-P_2;P_2)\,
S(p-P_2)\right]^{ab}
        + \int_k L^{ab;cd}(p+P_1,p-P_2;k-P_2,k+P1) \;
        \left[ \gamma_5 \; \gamma_A(P_2) \right]^{dc}
\label{pedro15a}
\end{eqnarray}
and similarly, for on-shell pions
\begin{eqnarray}
  \int_k L^{ab;cd}(p+P_1,p-P_2;k-P_2,k+P_1) \;
  \left[
\Gamma_A(k+P_1,k;P_1) \; S(k) \;
        \Gamma_\pi(k,k-P_2;P_2) \right]^{dc}
\nonumber \\  {} 
	= \left[\gamma_5\,S(p)\,\Gamma_\pi(p,p-P_2;P_2)\, 
	S(p-P_2)\right]^{ab} \, .
\label{pedro15pi}
\end{eqnarray}

%%%%%%%%%%%%%%%%%%%%%%%%%%%%%%%%%%%%%%%%%%%%%%%%%%%%%%%%%%%%%%%%%%
\subsection{$\pi -\pi$ scattering in rainbow-ladder truncation}

\begin{figure}[h]
\center
\epsfig{figure=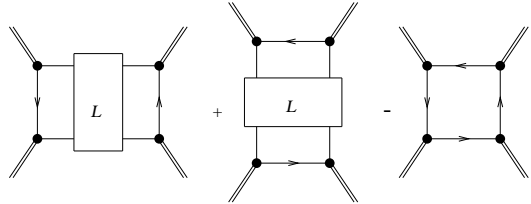,width=8cm}
\caption{
Amplitude for $\pi -\pi$ scattering in rainbow-ladder truncation that
reproduces Weinberg's result.  Note the $-$ sign for the direct
contribution due to the disconnected term in the amplitude $L$.}
\label{figamplitude}
\end{figure}

We now utilize Eq.~(\ref{expansion}) to evaluate the $\pi -\pi$
scattering amplitude near threshold.  Since there are six topologically
different ways to attach the external pion legs to the direct term,
we first calculate the contribution from one of these topologies by adding two sets of
diagrams with a complete set of ladder kernels $K$ inserted in the
direct contribution (see Fig.~\ref{figamplitude} and Ref.\cite{ladders}
for a additional details).
Then using the notation
introduced in the previous section yields the $\pi -\pi$
scattering amplitude, $A$ 
\begin{eqnarray}
 A \; = \; \int_k \int_p
        \left[ \bar{\Gamma}_\pi(p+P_3,p;P_3) \, S(p) \,
	\bar{\Gamma}_\pi(p,p-P_4;P_4) \right]^{ba} \;
	L^{ab;cd}(p-P_4,p+P_3;k-P_2,k+P_1) \; 
\nonumber\\  \times 
  	\left[ \Gamma_\pi(k+P_1,k;P_1) \, S(k) \,
        \Gamma_\pi(k,k-P_2;P_2) \right]^{dc}
\nonumber\\
  + \int_k \int_p
        \left[ \bar{\Gamma}_\pi(p+P_4,p;P_4) \, S(p)
	\Gamma_\pi(p,p-P_1;P_1) \right]^{dc} \;
	L^{ab;cd}(p-P_1,p+P_4;k+P_2,k-P_3) \;
\nonumber\\  \times
        \left[ \Gamma_\pi(k+P_2,k;P_2) \, S(k)
        \bar{\Gamma}_\pi(k,k-P_3;P_3) \right]^{dc}
\nonumber\\ 
  - \int_k {\rm Tr}[ \Gamma_\pi(k+P_1,k;P_1) \, S(k) \,
        \Gamma_\pi(k,k-P_2;P_2) \, S(k-P_2)
        \bar{\Gamma}_\pi(k-P_2,k-P_2-P_3;P_3) \, S(k-P_2-P_3) 
\nonumber\\  \times
	\bar{\Gamma}_\pi(k-P_2-P_3,k+P_1;P_4) \, S(k+P_1) ]
\end{eqnarray}
where the $P_i$ are constrained by momentum conservation,
$\sum_{i=1}^4 P_i=0$, and the on-shell condition, $P_i^2 = m_\pi^2$.
Note that there is a minus sign for the direct term because our
definition of $L$ includes the disconnected contribution (see
Eq.~(\ref{defL})).  Using the expansion Eq.~(\ref{expansion}) and the
relations Eqs.~(\ref{pedro36}) and (\ref{emilio15})-(\ref{pedro15pi}),
one can show that to order $P_i\,P_j$ the sum of these diagrams
reduces to
\begin{eqnarray}
	A &=& \frac{i}{4 f_\pi^2} 
	\left( (P_1+P_2)^2 + (P_1+P_4)^2 - 2 m_\pi^2 \right) \, .
\label{adler}
\end{eqnarray}
It immediately follows that the $\pi -\pi$ scattering amplitude at
threshold is proportional $( m_\pi/{f_\pi})^2$ and vanishes for
$m_\pi \rightarrow 0$.  This is the Adler zero.  Hence, in the chiral
limit the sum of the ladder diagrams exactly cancels the direct term
contribution.

For the physical scattering amplitude all six topologies must be
added, each with the appropriate combination of isospin factors.
In terms of the usual Mandelstam variables, $s$,
$t$ and $u$, the final result is
\begin{eqnarray}
  A_0 &=&  i \, { 2\,s - m_\pi^2 \over 2 f_\pi^2} \ ,
\end{eqnarray}
for the $I=0$ amplitude, and
\begin{eqnarray}
  A_2 &=& i \, { -\,s + 2 m_\pi^2 \over 2 f_\pi^2 } \ ,
\end{eqnarray}
for the $I=2$ amplitude, respectively.  This reduces to
the  Weinberg limit at threshold  ($s = 4 m^2_{\pi}$)
\begin{eqnarray}
 A_0 &=&  i \, {7 m_\pi^2 \over 2 f_\pi^2} = i \, 16 \, \pi \, a_0 \ ,
\end{eqnarray}
\begin{eqnarray}
 A_2 &=& -i \, { m_\pi^2 \over f_\pi^2} = i \, 16 \, \pi \, a_2  \ ,
\end{eqnarray}
independent of ladder kernel details. Here $a_I$ are the S wave
scattering lengths.  Hence, to properly describe $\pi -\pi$
scattering in the rainbow-ladder truncation, all possible diagrams
with one or more insertions of the ladder kernel $K$ must be combined
with the direct terms.  Note that for the direct terms there is
implicitly an infinite number of ladders inserted ``across one BSA in
a corner'' and on the bare quark lines.  Thus, for a consistent
calculation, the ladder kernel must be inserted, in {\em all} possible
ways, without crossing, in the skeleton diagrams like the ones in
Fig.~\ref{figdirect}.  This prescription can easily be generalized to
other processes even if they involve a different number of external particles.
In particular for processes with three external particles, the impulse
approximation in combination with the rainbow-ladder truncation for the
quark propagators and vertices does satisfy consistency requirements
following from current conservation (see Ref.\cite{Maris:2000bh} for an electromagnetic
application). In this case
all possible insertions
of the ladder kernel (without crossing) are already implicitly
included in the dressing of the propagators and vertices.

%%%%%%%%%%%%%%%%%%%%%%%%%%%%%%%%%%%%%%%%%%%%%%%%%%%%%%%%%%%%%%%%%%
\subsection{Numerical results at threshold}

\begin{figure}[h]
\center
\epsfig{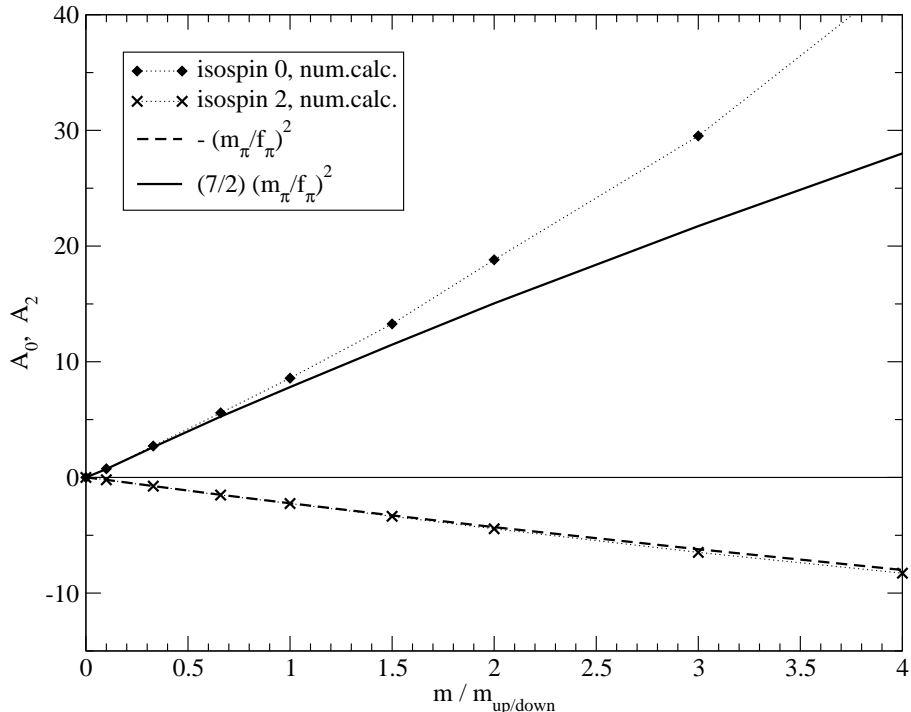}
\caption{ 
The $\pi -\pi$ scattering amplitudes at threshold calculated in
rainbow-ladder truncation as a function of the current quark mass.  For
comparison, we also show the Weinberg limit, using our calculated
values for $m_\pi$ and $f_\pi$.
\label{fig:pipires} }
\end{figure}

Utilizing the DSE model and calculation scheme of
Refs.~\cite{Maris:1999nt,Maris:2000bh}, we have conducted
a numerical analysis of $\pi -\pi$ scattering.
Figure~\ref{fig:pipires} shows our results for both isospin 
scattering amplitudes at threshold as a function of the current
quark mass (the imaginary phase has been suppressed).  The
corresponding pion mass and decay constant are also calculated as a
function of the quark mass and the square of their ratio is plotted as
well.  This model calculation clearly shows that the scattering
amplitudes indeed behave like $M^2_\pi/f_\pi^2$ if one includes the
two complete sets of ladder diagrams in addition to the direct term,
as indicated in Fig.~\ref{figamplitude}. In the chiral limit the two
sets of ladders cancel the direct contribution within the numerical
accuracy.

It is interesting that the direct term alone generates a scattering
amplitude about 20 times larger than observation.
For the physical value of the degenerate $u$, $d$ current quark mass,
$m_q = 5.54$ $MeV$ at renormalization scale $\mu = 1$ $GeV$, about 95\% of
the direct contribution is cancelled by including the ladder
diagrams. 
For this quark mass our corresponding scattering lengths are $a_0 =
0.170$ and $a_2 = -0.045$ which again are in excellent agreement with
Weinberg's values of $a_0 = 0.156$ and $a_2 = -0.044$ (using $m_{\pi}
= 138.04 \,{\rm MeV}$ and $f_{\pi} = 92.44 \,~{\rm MeV}$). This is
also in good agreement with the physical $\pi -\pi$ scattering
lengths, $a_0 = 0.220$ and $a_2 = -0.0444$.

It is also interesting to note that chiral perturbation theory 
 reproduces even more precisely the physical scattering lengths \cite{col}.  This
is due to pion loops, which we have not included, and the isospin $0$
channel is more sensitive than the isospin $2$ channel to this effect.
The 3rd order chiral perturbation theory for $a_2$ is almost 
identical to Weinberg's value, whereas there are significant
corrections from 2nd and even 3rd order chiral perturbation
theory to Weinberg's $a_0$.

In addition to chiral symmetry there is another important correction that is 
implicitly included in our
approach. This is the effect from scalar bound states.  In particular an idealized $\sigma$
meson, which appears as a $q\bar{q}$ bound state in our model, is present and also
influences the isospin zero scattering amplitude.

%%%%%%%%%%%%%%%%%%%%%%%%%%%%%%%%%%%%%%%%%%%%%%%%%%%%%%%%%
%%%%%%%%%%%%%%%%%%%%%%%%%%%%%%%%%%%%%%%%%%%%%%%%%%%%%%%%%%%%%%%%%%%%%%%%%%%%%
\section{Further comments and conclusion}
\label{secCon}
Summarizing, we find  chiral symmetry preserving quark models can indeed yield the correct
$I=0$ and $I=2$ $\pi -\pi$ scattering amplitudes near
threshold, $(2s- m_\pi^2) /( 2 i\, f_\pi^2)$ and $ (-s+2 m_\pi^2) /
(2 i\, f_\pi^2)$, respectively.  Further, we have rigorously reproduced Weinberg's 
low energy theorem
for the scattering
lengths.  We have
proved this independent of the interaction details for   a manifestly covariant DSE/BSE approach
using the rainbow-ladder truncation, and also for an instantaneous Hamiltonian formulation in the
infinite interaction limit.  In
the former case this result has also been demonstrated by summing the
series of diagrams in Fig.~\ref{figamplitude} numerically using the
model of Ref.~\cite{Maris:1999nt,Maris:2000bh}. Lastly, both approaches produce the Adler 
zero which provides new and 
important bounds for
couplings to scalar mesons. 

We expect our result to be a general feature in any quark model  preserving
both chiral symmetry and the AV-WTI. However, care is
necessary to ensure  any truncation does indeed satisfy {\em
all} chiral symmetry constraints.  For ladder truncations, the
crucial step is to include all diagrams with the ladder kernel inserted
in the direct contribution.  Finally, we submit this
prescription is also necessary for consistency in other processes
(e.g. current
conservation),  as well as
reactions involving more than four external
particles.

%%%%%%%%%%%%%%%%%%%%%%%%%%%%%%%%%%%%%%%%%%%%%%%%%%%%%%%%%%%%%%%%%%%%%%%%%%%%%
\acknowledgements

Pedro Bicudo acknowledges Gon\c{c}alo Marques' assistance and enlightening discussions with Gastao
Krein, George Rupp and Mike Scadron.  This work is supported in part by grants DOE DE-FG02-97ER41048
and NSF INT-9807009.

%%%%%%%%%%%%%%%%%%%%%%%%%%%%%%%%%%%%%%%%%%%%%%%%%%%%%%%%%%%%%%%%%%%%%%%%%%%%%
%\section*{References}
%

%%%%%%%%%%%%%%%%%%%%%%%%%%%%%%%%%%%%%%%%%%%%%%%%%%%%%%%%%%%%%%%%%%%%%%%%%%%%%
%
\end{document}